\documentclass[onecolumn,pra,aps,showpacs,reprint]{revtex4-1} 
\usepackage{amsmath,mathrsfs,amsbsy,color,graphicx,bm,amsthm,amsfonts}
\usepackage{times}
\usepackage{graphicx}
\usepackage{braket}
\usepackage{dsfont}
\usepackage{hyperref}       
\usepackage{url}            
\usepackage{booktabs}       
\usepackage{nicefrac}       
\usepackage{microtype}      
\usepackage{lipsum}
\usepackage{float}
\usepackage{tikz}
\usetikzlibrary{decorations.pathreplacing,matrix}

\newtheorem{theorem}{Theorem}

\begin{document}
	\title{Wigner's friend and the quasi-ideal clock}

	\author{Vinicius P. Rossi}
	\email{prettirossi.vinicius@gmail.com}
	\author{Diogo O. Soares-Pinto}
	\email{dosp@usp.br}
	\affiliation{Instituto de Física de São Carlos, Universidade de São Paulo, CP 369, 13560-970, São Carlos, São Paulo, Brazil}
	
	\begin{abstract}
		In 1962, Eugene P. Wigner introduced a thought experiment that highlighted the incompatibility in quantum theory between unitary evolution and wave function reduction in a measurement. This work resulted in a class of thought experiments often called Wigner's Friend Scenarios, which have been providing insights over many frameworks and interpretations of quantum theory. Recently, a no-go theorem obtained by Daniela Frauchiger and Renato Renner brought attention back to the Wigner’s Friend and its potential of putting theories to test. Many answers to this result pointed out how timing in the thought experiment could be yielding a paradox. In this work, we ask what would happen if the isolated friend in a Wigner's Friend Scenario did not share a time reference frame with the outer observer, and time should be tracked by a quantum clock. For this purpose, we recollect concepts provided by the theory of quantum reference frames and the quantum resource theory of asymmetry, to learn how to internalize time in this scenario, and introduce a model for a feasible quantum clock proposed by Mischa P. Woods, Ralph Silva and Jonathan Oppenheim, called the quasi-ideal clock. Our results have shown that no decoherent behavior comes from this approach, and the disagreement between the superobserver and its friend persists even for an imprecise clock on Wigner's side. However, the gaussian spread of this clock model can control what observables do not raise a paradox, indicating the relevance of deepening this analysis.
	\end{abstract}
	
	\maketitle
	\section{Introduction}
	
	Dynamics in quantum theory is given by two well-known postulates: one of them, the Schr\"odinger equation, describes the time evolution of isolated systems via unitary operators, while the second one, the measurement postulate, describes how a system is to be described after interacting with a measurement apparatus. To emphasize the incompatibility between these two descriptions, Eugene P. Wigner proposed a thought experiment later called the Wigner's friend \cite{Wigner1995}, in which an observer inside a lab would measure a physical binary property of a quantum system, this being followed by a second global measurement of the whole lab, made by an external superobserver. Given the right initial state of the quantum system and a set of measurements for the observers, they cannot agree on the probability distribution of the external observer's outcomes.
	
	What is indeed happening in this protocol is the following: with respect to an ideal classical clock, the internal friend, that from now on should be called Alice, performs her measurement over the quantum system. The superobserver, called Wigner in this work, measures the same ideal classical clock, or a synchronized and perfect copy of it, and when he is sure Alice is done with her measurement, he performs his own one. It sounds unreasonable, however, to quantum mechanically describe a complex system such as a lab (which includes every degree of freedom of Alice and her measurement device) and not to do so with the clock.
	
	In 2018, Daniela Frauchiger and Renato Renner \cite{Frauchiger2018} published an article pointing out how Wigner's friend scenarios (WFS) could do more than highlight this fundamental incompatibility: they can be used as a test environment for the compatibility between assumptions about the world. Their extended WFS resulted in a no-go theorem stating that ($\mathcal{Q}$) the universal validity of quantum theory, ($\mathcal{C}$) the consistency between predictions made by different agents and ($\mathcal{S}$) single-world interpretations do not agree between themselves and shall lead to a paradox if simultaneously imposed over this scenario. Many responding articles argued, however, how there were other hidden assumptions that could be producing the paradox instead of ($\mathcal{Q}$), ($\mathcal{C}$) and ($\mathcal{S}$), and the authors themselves claim that the definition of concepts such as \emph{time} may be a source of paradox instead.
	
	In this work, we address to the question of what would happen if we described the clock with respect to which Wigner performs his measurement as a quantum system. By imposing this condition, Wigner and Alice should no longer share a clock, since Alice's lab would cease to be isolated, and Schr\"odinger equation would not apply. Wigner will not know when Alice's measurement is done, but will know by a common established protocol when he is supposed to measure the lab state. Furthermore, with the aim of inserting a source of uncertainty that could produce the desired decoherence, we shall equip Wigner with a specific quantum clock, proposed by Woods, Silva and Oppenheim in 2019 \cite{Woods2019}.
	
	This work is structured as follows: Section \ref{sec:2} will revise the Wigner's friend scenarios of Wigner and Frauchiger-Renner, and sketch our own WFS. Section \ref{sec:3} briefly review the problem of marking time in quantum mechanics, calling upon the theory of quantum reference frames to provide us the necessary tools for our model, and also introducing the Woods-Silva-Oppenheim clock states. Section \ref{sec:4} will finally propose a model for the lab dynamics, and Section \ref{sec:5} will derive our results, mainly stating that for the proposed models of lab and clock evolutions, the external observer still disagrees with the predictions of the internal agent for most of his possible measurement choices. We also analyze how the external agent's measurements are asymmetric with respect to time evolution. Finally, Section \ref{sec:6} concludes with suggestions for further work.
	\section{Wigner's friend scenarios}\label{sec:2}
	
	The simplest Wigner's friend scenario can be described as the scheme given by Fig. \ref{fig:1}. Alice, inside her lab, is going to measure $\sigma_z$ over and ensemble of spin-$\frac{1}{2}$ particles. Let us assume that the ensemble is described by the quantum state
	\begin{equation}
		\ket{\psi}=\frac{1}{\sqrt 2}(\ket{\uparrow}_S+\ket{\downarrow}_S).
	\end{equation}
	Thus after performing a selective measurement, started at time $t_0$ and finished at time $t_A$ with respect to the shared  classical clock, she should describe the lab state as either
	\begin{equation}
		\ket{\Psi_+}=\ket{\uparrow}_S\otimes\ket{\uparrow}_A, \quad \mbox{or} \quad \ket{\Psi_-}=\ket{\downarrow}_S\otimes\ket{\downarrow}_A,
	\end{equation}
	where $\{\ket{\uparrow}_A,\ket{\downarrow}_A\}$ represent the state of Alice's measurement device, body and mind, and what else might exist inside the lab and might change with the measurement.
	
	Let us assume now that Wigner is going to perform a projection of the lab over the space associated to the state
	\begin{equation}
		\ket{ok}=\frac{1}{\sqrt 2}(\ket{\uparrow}_S\otimes\ket{\uparrow}_A-\ket{\downarrow}_S\otimes\ket{\downarrow}_A),
	\end{equation}
	registering the outcome $ok$ if this projection is successful, and $fail$ otherwise. From Alice's perspective thus Wigner is going to observe his outcomes at time $t_W$ with probabilities
	\begin{equation}
		P_A(ok|\uparrow)=|\braket{ok|\Psi_+}|^2=\frac{1}{2};
	\end{equation}
	\begin{equation}
		P_A(ok|\downarrow)=|\braket{ok|\Psi_-}|^2=\frac{1}{2}.
	\end{equation}
	From Wigner's perspective, however, there is no state-reduction inside the lab, since it is isolated and can only evolve unitarily. Alice's measurement results for Wigner in a state at time $t_A$ given by
	\begin{equation}
		\ket{\Phi_+}=\frac{1}{\sqrt 2}(\ket{\uparrow}_S\otimes\ket{\uparrow}_A+\ket{\downarrow}_S\otimes\ket{\downarrow}_A),
	\end{equation}
	which is orthogonal to the projection Wigner aims to detect. Therefore, Wigner predicts his probability distribution at time $t_W$ to be
	\begin{equation}
		P_W(ok)=|\braket{ok|\Phi_+}|^2=0;
	\end{equation}
	\begin{equation}
		P_W(fail)=1-P_W(ok)=1.
	\end{equation}
	
	The paradox will only vanish, i.e., Alice and Wigner will predict the same probability distribution only for 
	\begin{equation}\label{wigner0}
		\ket{ok}=\frac{1}{\sqrt 2}(\ket{\uparrow}_S\otimes\ket{\uparrow}_A\pm i\ket{\downarrow}_S\otimes\ket{\downarrow}_A).
	\end{equation}
	
	\begin{widetext}
		\begin{center}
			\begin{figure}[hbt!]
				\centering
				\includegraphics[scale=0.6]{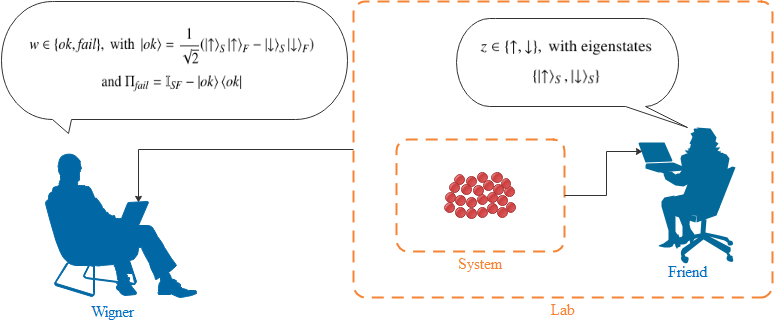}
				\caption{Schematic model of the simplest Wigner's friend scenario.}
				\label{fig:1}
			\end{figure}
		\end{center}
	\end{widetext} 
	
	Frauchiger and Renner propose a more complex version of this experiment, constituted of two labs. The protocol goes as follows, with $n$ being the steps:
	\begin{itemize}
		\item $n=00$: Alice, inside her lab, measures the side of a quantum coin, given by the value $r\in\{h,t\}$. The coin is prepared in the state
		\begin{equation}
			\ket{\psi}_C=\frac{1}{\sqrt 3}\ket{h}_C+\sqrt{\frac{2}{3}}\ket{t}_C.
		\end{equation}
		If she gets $r=h$, she prepares a spin-$\frac{1}{2}$ particle in the state $\ket{\downarrow}_S$, and if she gets $r=t$, she prepares it in the state $\frac{1}{\sqrt 2}(\ket{\uparrow}_S+\ket{\downarrow}_S)$. She then sends this spin-$\frac{1}{2}$ particle through a quantum channel to the neighbor lab.
		\item $n=10$: Bob, inside the neighbor lab, is going to detect the spin $z\in\left\{+\frac{1}{2},-\frac{1}{2}\right\}$, and nothing more.
		\item $n=20$: After all measurements inside the labs were carried out, the external observer Ursula is going to perform a projection over Alice's lab with respect to the state
		\begin{equation}
			\ket{ok}_U=\frac{1}{\sqrt 2}(\ket{h}_C\otimes\ket{h}_A-\ket{t}_C\otimes\ket{t}_A),
		\end{equation}
		where $\{\ket{h}_A,\ket{t}_A\}$ represent Alice's device, body and mind, just like in the previous WFS. She registers $u=ok$ if the projection is successful, and $u=fail$ otherwise.
		\item $n=30$: Another superobserver, Wigner, will do the same over Bob's lab, projecting it with respect to the state
		\begin{equation}
			\ket{ok}_W=\frac{1}{\sqrt 2}(\ket{\downarrow}_S\otimes\ket{\downarrow}_B-\ket{\uparrow}_S\otimes\ket{\uparrow}_B),
		\end{equation}
		where $\{\ket{\uparrow}_B,\ket{\downarrow}_B\}$ represent Bob's devide, body and mind. He registers $w=ok$ if the projection is successful, and $w=fail$ otherwise.
		\item $n=40$: If $u=ok$ and $w=ok$, the experiment is halted. Otherwise, it is reset.
	\end{itemize}
	
	Furthermore, every agent in this extended WFS shares three reasonable assumptions about the world:
	\begin{itemize}
		\item \emph{Universal validity of quantum theory} ($\mathcal{Q}$): any system can be correctly described by a state $\ket{\psi}$ in a Hilbert space, and its physical properties are given by projections of this state with respect to a family of Heisenberg projectors defined in a given time $t_0$, $\{\Pi_x(t_0)\}_{x\in\chi}$, being completed at time $t\geq t_0$. If $\braket{\psi|\Pi_\xi(t_0)|\psi}=1$, then an agent can properly says that \emph{``I know that $x=\xi$ at the time $t$"}.
		\item \emph{Self-consistency} ($\mathcal{C}$): If agents A and B reason over the same theory, and agent A can state that \emph{``I know that agent B knows that $x=\xi$ at time $t$"}, then she can also say \emph{``I know that $x=\xi$ at time $t$"}.
		\item \emph{Single-world} ($\mathcal{S}$): Physical quantities can have only one value at a given time $t$. In other words, if an agent can say that \emph{``I know that $x=\xi$ at time $t$"}, then he must \emph{deny} that \emph{``I know that $x\neq\xi$ at time $t$"}.
	\end{itemize}
	It is important to emphasize that assumptions ($\mathcal{Q}$) and ($\mathcal{C}$) were explicitly assumed by Wigner in his original work, while assumption ($\mathcal{S}$) was implicitly assumed.
	
	From the perspective of Ursula and Wigner, every measurement inside the lab is described as a von Neumann measurement, and is not hard to see that, at step $n=21$, when every internal measurement is done, the global state yields to a probability distribution such that
	\begin{equation}
		P(w=ok,u=ok)=\frac{1}{12}.
	\end{equation}
	However, even though Wigner, Ursula and Bob, using assumptions ($\mathcal{Q}$) and ($\mathcal{C}$) all agree that Alice must detect $r=t$ at step $n=01$ for the halting condition to be satisfied, when we assume that Alice indeed measured $r=t$ from her perspective, she will conclude that $w=fail$ at step $n=31$, which must be false by assumption ($\mathcal{S}$).
	
	We see by this arguing how the time marking is relevant and can be confusing in this sort of thought experiment. Many works in literature pointed out how this timing might be generating the paradox instead of assumptions ($\mathcal{Q}$), ($\mathcal{C}$) and ($\mathcal{S}$). Sudbery \cite{Sudbery2019} lists it among many other hidden assumptions in Frauchiger and Renner's work. Losada, Laura and Lombardi \cite{Lombardi2019} analyse the extended WFS under the consistent stories interpretation, to conclude that this sequence of statements do not belong to the same consistent chain of events. Waaijer and Van Neerven \cite{Waaijer2019} pointed out how agents' statements rely on registers from the past that are not in fact happening, which is forbidden for relational quantum mechanics. Baumann et al. \cite{Baumann2019} include a quantum ideal clock in a WFS, deriving some conditional probabilities that might rule out the paradox, and finally Gambini, García-Pintos and Pullin \cite{Gambini2019} propose that uncertainties in time and length measurements are fundamental to ensure the indistinguishability between a reduced state and a decohered one, claiming that this might solve the so-called Frauchiger-Renner paradox. These are few examples of how time might play a crucial role on solving the Wigner's friend problem.
	
	We here argue that one questionable feature of the Wigner's friend scenario is that Alice and Bob are in fact sharing a time (classical) reference frame (Fig. \ref{fig:2}a). If we are going to assume by (Q) that \emph{any} physical system is going to be described by a quantum state, there is no reason why one should not describe the clock by a vector in a Hilbert space of its own. This being made, we argue that the agents can no longer share a clock, because if they did so, then Alice's lab would be an open quantum system (Fig. \ref{fig:2}b). From now on, we will privilege Wigner's perspective \cite{Zukowski2020}, even though it should not make any difference.
	
	\begin{widetext}
		\begin{center}
			\begin{figure}[hbt!]
				\includegraphics[scale=0.5]{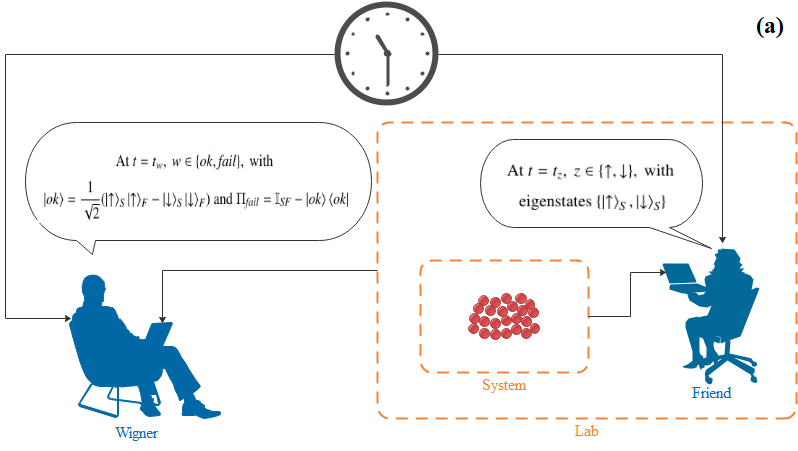}\\
				\includegraphics[scale=0.54]{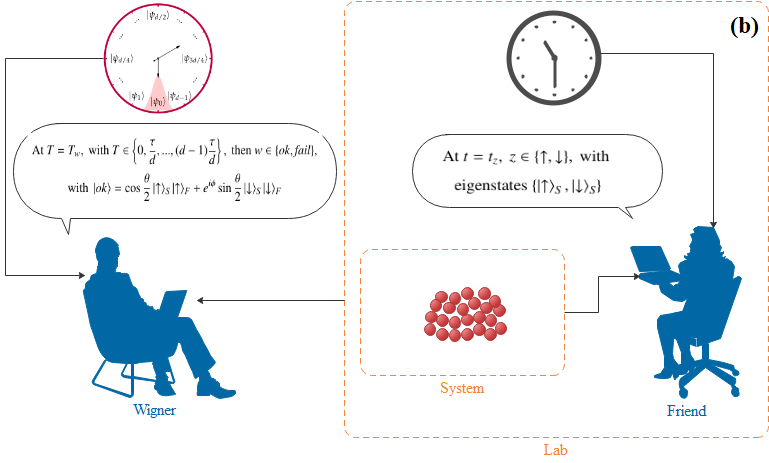}
				\caption{Schematic representation of the simplest Wigner's friend scenario, including explicitly the time reference frame for two scenarios: (a) Alice and Wigner share a classical clock; (b) Alice and Wigner have their own clocks.}
				\label{fig:2}
			\end{figure}
		\end{center}
	\end{widetext}
	
	\section{Quantum time and quantum clocks} \label{sec:3}
	
	To build a quantum operator capable of telling what time is it, we expect it has some specific properties, such as \cite{Pashby2014}
	\begin{equation}\label{cov}
		U_t^\dagger TU_t=T+t,
	\end{equation}
	where $T$ is the time operator in the Schr\"odinger picture, $U_t$ is a representation of an element of the uniparametric strongly continuous group generated by an hamiltonianian $H$, and $t$ is the parameter of the Schr\"odinger equation. Eq. (\ref{cov}) is typically known as the \emph{global covariance relation}. This immediately lead to the canonical commutation relation
	\begin{equation}
		[T,H]=i.
	\end{equation}
	However, Wolfgang Pauli proposed an argument \cite{Galapon2002} that threw pessimism over the construction of a time operator. This argument is hereby introduced as a theorem, and goes as follows:
	\begin{theorem}
		\emph{(Pauli)} Let $\mathcal{H}$ be a separable Hilbert space, and let $H,T\in\mathcal{B}(\mathcal{H})$ be self-adjoint operators acting on this Hilbert space. Then, if $T$ obeys a global covariance relation with each element of the uniparametric strongly continuous group of unitaries generated by $H$, i.e., if
		\begin{equation}\label{cov2}
			U_t^\dagger TU_t=T+t, \quad \forall t\in\mathbb{R},
		\end{equation}
		then the spectra $\mbox{spec}(H)$ and $\mbox{spec}(T)$ are both equivalent to $\mathbb{R}$.
	\end{theorem}
	
	This was taken as a result that forbids the existence of a time operator, because it would take a hamiltonian unbounded from bellow to recover the global covariance relation, which is not allowed by thermodynamics. However, it is important to highlight that the theorem does not rule out every possibility of building a time operator. It just states that (i) a time operator $T$ with spectrum equivalent to $\mathbb{R}$ and (ii) a hamiltonian $H$ bounded from bellow cannot be related to each other by Eq. (\ref{cov2}). 
	
	
	\subsection{Page-Wootters Mechanism}\label{sec:3.1}
	
	The first step to solve this problem was taken by Paul Dirac in 1926 \cite{Dirac1926}, in a procedure of extending the Hilbert space that would later be used by Bryce DeWitt in the construction of his constraint equation for quantum gravity \cite{DeWitt1967}. What interest us is the solution proposed by Don Page and William Wootters in 1983 \cite{Page1983,Wootters1984}. It consists of a universe described by a bipartite Hilbert space, $\mathcal{H}=\mathcal{H}_A\otimes\mathcal{H}_B$, whose dynamics is governed by the non-interacting hamiltonian
	\begin{equation}\label{global}
		H=H_A\otimes\mathbb{I}_B+\mathbb{I}_A\otimes H_B.
	\end{equation}
	In the Page-Wootters mechanism (PaW), the only states truly accessible for an observer are solutions of the constraint equation
	\begin{equation}\label{PaW}
		H\ket{\Psi}\rangle=0.
	\end{equation}
	That is because, in this universe, the parametric time $t$ in the Schr\"odinger equation is inaccessible. Instead, the dynamics of the subsystem $A$ is given \emph{relationally} with respect to subsystem $B$. If there is a way of building a time operator $T_B$, with $\mbox{spec}(T_B)\equiv\mathbb{R}$, $[T_B,H_B]=i$ and non degenerate eigenvectors $\{\ket{\phi_B(t)}\}_{t\in\mbox{spec}(T_B)}$, we are allowed to describe the local state of system $A$ as
	\begin{equation}
		\ket{\psi_A(t)}=\frac{\braket{\phi_B(t)|\Psi}\rangle}{|\braket{\phi_B(t)|\Psi}\rangle|},
	\end{equation}
	and it is even possible to show that Schr\"odinger equation is recovered on $A$, i.e.,
	\begin{equation}
		i\frac{d}{dt}\ket{\psi_A(t)}=H_A\ket{\psi_A(t)},
	\end{equation}
	except that now $t$ is no classical parameter, but rather an eigenvalue of an operator. Once the global physical state $\ket{\Psi}\rangle$ is known, the dynamics in system $A$ can thus be derived from it. This leads, however, to two important questions:
	\begin{enumerate}
		\item Typically, an agent has no access to this physical state. Instead, what is known is a prepared state $\rho$, that in this work will always be a product state, and whose evolution still depends on the Schr\"odinger equation parameter $t$. How does one start from $\rho$ and obtain $\ket{\Psi}\rangle$, from which the relational description is to be derived?
		\item There is still a local problem in subsystem $B$, since our time operator might be fulfilling every condition to be ruled out by Pauli's argument. Is there a physical system capable of emulating every property of a quantum ideal clock, but that is still described by a bounded hamiltonian?
	\end{enumerate}
	We aim to answer to these questions in the following sections.
	
	\subsection{Internalizing time}\label{sec:3.2}
	
	To answer to the first question, we call upon the theory of quantum reference frames \cite{Bartlett2007}. It deals with problems where two parties, Alice and Bob, with their own quantum systems, described with respect to their own quantum reference frames, communicate with each other in the absence of a shared classical reference frame. To illustrate this sort of problem, we can think of Alice and Bob scheduling a date for 2 p.m. at the park. However, Bob has just arrived from a distant country, and has no idea of which timezone they are in. He has his watch with him, but the lack of a classical reference frame between them made it almost useless. In this situation, what should Bob do? To ensure that Alice will not be left alone waiting for him, he could just go to the park as soon as he can, and sit on a bench until Alice shows up. What Bob is in fact doing is an average over every possible reference frame that might exist between him and Alice. In the theory of quantum reference frames, this operation is known as \emph{$G$-twirling}, and is given by
	\begin{equation}
		\mathcal{G}[\rho]=\int_{g\in G}U_g\rho U_g^\dagger dg,
	\end{equation}
	where $g\in G$ is an element of the group of transformations between reference frames in question, $U_g$ is its representation and $dg$ is the Haar measure \cite{Barut1980,Carmo2020}.
	
	It is easy to show that, if $G$ is the group of time translations described by the global hamiltonian in Eq. (\ref{global}), then $\mathcal{G}[\rho]$ is the static solution of the dynamical equation of motion in the density operator formalism. Indeed, for $\frac{\partial H_S}{\partial t}=\frac{\partial H_C}{\partial t}=0$ and $\frac{\partial\rho}{\partial t}=0$ in the Schrödinger picture, then
	\begin{equation}
		\left[\mathcal{G}[\rho],H\right] = \lim_{T\rightarrow\infty}\frac{1}{2T}\int_{-T}^TU_t[\rho,H]U_t^\dagger dt,
	\end{equation}
	and since for these states $\frac{d\rho}{dt}=iU_t[\rho,H]U_t^\dagger$, then
	\begin{equation}
		\left[\mathcal{G}[\rho],H\right]=-i\lim_{T\rightarrow\infty}\frac{\rho(T)-\rho(-T)}{2T}=0,
	\end{equation}
	for $\rho(T)$ and $\rho(-T)$ have both finite eigenvalues for every $T$. The relative state of the subsystem $S$ is thus written as
	\begin{equation}\label{relative}
		\rho_S(t)=\frac{\mbox{Tr}_C\{(\mathbb{I}_S\otimes\Pi_t^C)\mathcal{G}[\rho](\mathbb{I}_S\otimes\Pi_t^C)\}}{\mbox{Tr}\{(\mathbb{I}_S\otimes\Pi_t^C)\mathcal{G}[\rho]\}},
	\end{equation}
	where $\Pi_t^C$ are projectors over eigenspaces associated to the eigenvectors $\ket{\phi_C(t)}$ of $T$. A typical example is to think of an universe made of two qubits. Their initial states will be both described as $\rho=\ket{+}\bra{+}$, and the non-interacting hamiltonian is given by
	\begin{equation}
		H=\omega(\sigma_z^S\otimes\mathbb{I}_C+\mathbb{I}_S\otimes\sigma_z^C).
	\end{equation}
	The $G$-twirling operation over the group generated by this hamiltonian gives us the symmetrized state (in the computational basis for two qubits)
	\begin{equation}
		\mathcal{G}[\rho]=\frac{1}{4}
		\begin{pmatrix}
			1 & 0 & 0 & 0\\
			0 & 1 & 1 & 0\\
			0 & 1 & 1 & 0\\
			0 & 0 & 0 & 1
		\end{pmatrix}.
	\end{equation}
	If we now define the time operator over the clock system $C$ to be $T_C=\sigma_x^C$, in the sense that eigenvalues of this operator are shifted by the unitary $U_t$ in discreet steps $t=\pi/\omega$, i.e.,
	\begin{equation}
		U_{\pi/\omega}\ket{\pm}=\ket{\mp},
	\end{equation}
	then we can obtain the relative state of system $S$ as being
	\begin{equation}
		\rho_S(\pm)=\frac14
		\begin{pmatrix}
			2 & \pm1\\
			\pm1 & 2
		\end{pmatrix}.
	\end{equation}
	For this state, the probabilities of agreement between the system and the clock are $P(+|+)=P(-|-)=\frac{3}{4}$, while there will be a mistracking with probability $P(+|-)=P(-|+)=\frac{1}{4}$. This happens because the pair $T_C,H_C$ does not constitute a canonical pair, and thus this qubit is not the best choice for a quantum clock. 
	
	
	\subsection{Quasi-ideal clock states}\label{sec:3.3}
	
	To answer the second question from Sec. \ref{sec:3.1}, we turn ourselves to the model of quantum clock proposed by A. Peres in 1980 \cite{Peres1980}. Known as the Salecker-Wigner-Peres clock, due to the pioneer work of H. Salecker and Wigner himself on the formulation of a quantum clock in 1958 \cite{Salecker1958}, it consists of a finite dimensional system described by the hamiltonian
	\begin{equation}\label{HC}
		H_C=\omega\sum_{n=0}^{d-1}n\ket{n}\bra{n}.
	\end{equation}
	In order to extract the canonical commutation $[T_C,H_C]=i$, we want to build a basis in which $T_C$ is diagonal and that is related to the energy basis in the same way as the momentum basis is related to position basis, since momentum and position operators are canonically conjugated. This invites us to write the basis
	\begin{equation}
		\ket{\theta_k}=\frac{1}{\sqrt{d}}\sum_{n=0}^{d-1}e^{-i2\pi nk/d}\ket{n},
	\end{equation}
	with $k\in\{0,d-1\}\subset\mathbb{Z}$. This basis has interesting properties, such as the discreet shift in steps $t=\frac{2\pi}{\omega d}$, i.e.,
	\begin{equation}
		U_{\tau/d}\ket{\theta_k}=\ket{\theta_{k+1}},
	\end{equation}
	and each vector $\ket{\theta_k}$ is infinitely degenerated,
	\begin{equation}
		\ket{\theta_k}=\ket{\theta_{k+md}}, \quad m\in\mathbb{Z}.
	\end{equation}
	Building a time operator diagonal in this basis,
	\begin{equation}\label{T}
		T_C=\frac{\tau}{d}\sum_{k=0}^{d-1}k\ket{\theta_k}\bra{\theta_k},
	\end{equation}
	it is expected that the canonical commuting relation should be obtained. However, it is easy to see that
	\begin{equation}
		\braket{\theta_k|[T_C,H_C]|\theta_k}=0, \quad \forall\,k\in\{0,d-1\}.
	\end{equation}
	However, Peres applied this model of clock in three classic problems of quantum mechanics, and showed that this system could indeed keep track of dynamics with an imprecision due to the evolution in discreet steps. He also argued that any attempt to increase the clock dimension in order to improve its precision would eventually lead to interaction between the system and the clock, affecting (and eliminating, in many cases) the observed phenomenon.
	
	The improve in this clock model was recently given by M. Woods, R. Silva and J. Oppenheim in 2019, with their model of quasi-ideal clock states \cite{Woods2019}. It consists of a Salecker-Wigner-Peres clock, with the states of interest being no longer the pointer states, but rather a superposition of them, given by
	\begin{equation}
		\ket{\psi(k_0)}=\sum_{k\in S_d(k_0)}Ae^{-\frac{\pi}{\sigma^2}(k-k_0)^2}e^{i2\pi n_0(k-k_0)/d}\ket{\theta_k},
	\end{equation}
	where $k_0$ is a real number related to the parameter in the Schr\"odinger equation, $k_0=td/\tau$, $\sigma$ is a gaussian standard deviation and $n_0$ is associated to the mean energy of the state, $\braket{H}_{\psi(k_0)}=n_0\omega$. $A$ is a normalization factor, and $S_d(k_0)$ is given by
	\begin{equation}
		S_d(k_0):=
		\begin{cases}
			\left\{k\in\mathbb{Z}|-\frac{d}{2}\leq k-k_0<\frac{d}{2}\right\},\mbox{ even }d \\
			\left\{k\in\mathbb{Z}+\frac12|-\frac{d}{2}\leq k-k_0<\frac{d}{2}\right\},\mbox{ odd }d
		\end{cases}.
	\end{equation}
	It is very interesting to see that the expectation value of the time operator given by Eq. (\ref{T}) for these states covaries with the external time $t$, as it can be seen in Fig. \ref{Fig:3}. There are two relevant results for these quasi-ideal states that we enunciate bellow:
	\begin{theorem}
		\emph{(Quasi-continuity)} Let $\mathcal{H}$ be the Hilbert space of a Salecker-Wigner-Peres clock, with $H$ being its hamiltonian and $T$ being the time operator. Let $\ket{\psi(k_0)}$ be a quasi-ideal clock state. Then, for any $t\in\mathbb{R}$,
	\end{theorem}
	
	\begin{widetext}
		\begin{equation}
			e^{-iH_Ct}\ket{\psi(k_0)}=\sum_{k\in S_d(k_0+td/\tau)}Ae^{-\frac{\pi}{\sigma^2}(k-k_0+td/\tau)^2}e^{i2\pi n_0(k-k_0+td/\tau)}\ket{\theta_k}+\ket{\epsilon},
		\end{equation}
	\end{widetext}
	
	\textit{with}
	\begin{equation}
		|\braket{\theta_k|\epsilon}|\leq\mathcal{O}\left(t\,\mbox{poly}(d)e^{-\frac{\pi d}{4}}\right), \quad d\rightarrow\infty.
	\end{equation}
	
	This means that, for a clock size that is large enough, the hamiltonian $H_C$ generates a continuous shift in the quasi-ideal clock states up to a vanishing error. 
	
	\begin{theorem}
		\emph{(Quasi-canonical commutation)} Let $\mathcal{H}$ be a Hilbert space of a Salecker-Wigner-Peres clock, with $H_C$, $T_C$, and $\ket{\psi(k_0)}$ be the previously defined operators and state. Then,
		\begin{equation}
			[T_C,H_C]\ket{\psi(k_0)}=i\ket{\psi(k_0)}+\ket{\epsilon_c},
		\end{equation}
		with 
		\begin{equation}
			|\braket{\epsilon_c|\epsilon_c}|^2\leq\mathcal{O}\left(\mbox{poly}(d)e^{-\frac{\pi d}{4}}\right), \quad d\rightarrow\infty.
		\end{equation}
	\end{theorem}
	With this theorem, the statistics of a canonical pair is recovered for quasi-ideal clock states, evading the Pauli's argument and still keeping track of time in a satisfactory way. These results must be enough to ensure the applicability of the Woods-Silva-Oppenheim clock in our model of WFS.
	
	\begin{widetext}
		\begin{center}
			\begin{figure}[hbt!]
				\centering
				\begin{tabular}{cc}
					\includegraphics[scale=0.5]{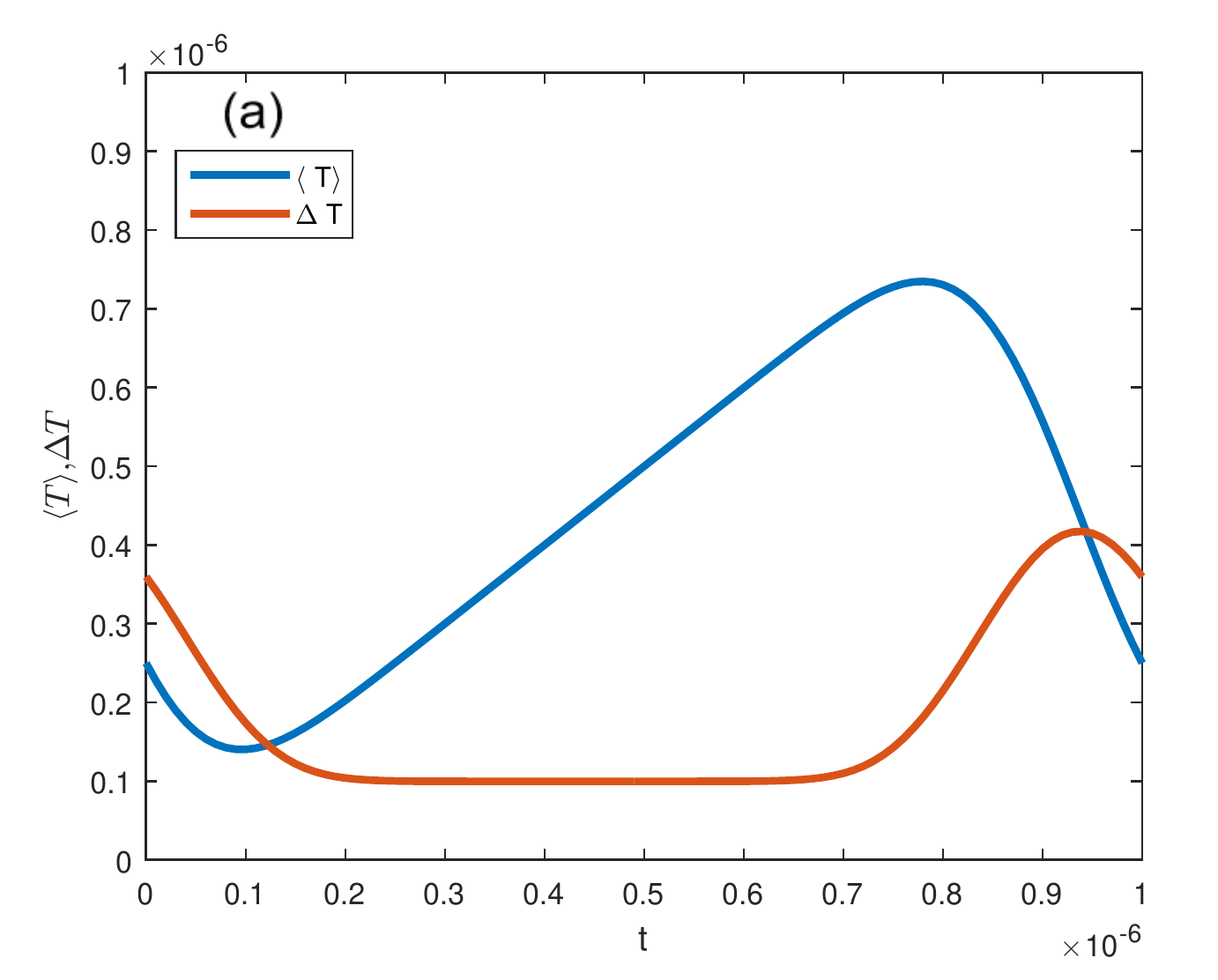} & \includegraphics[scale=0.5]{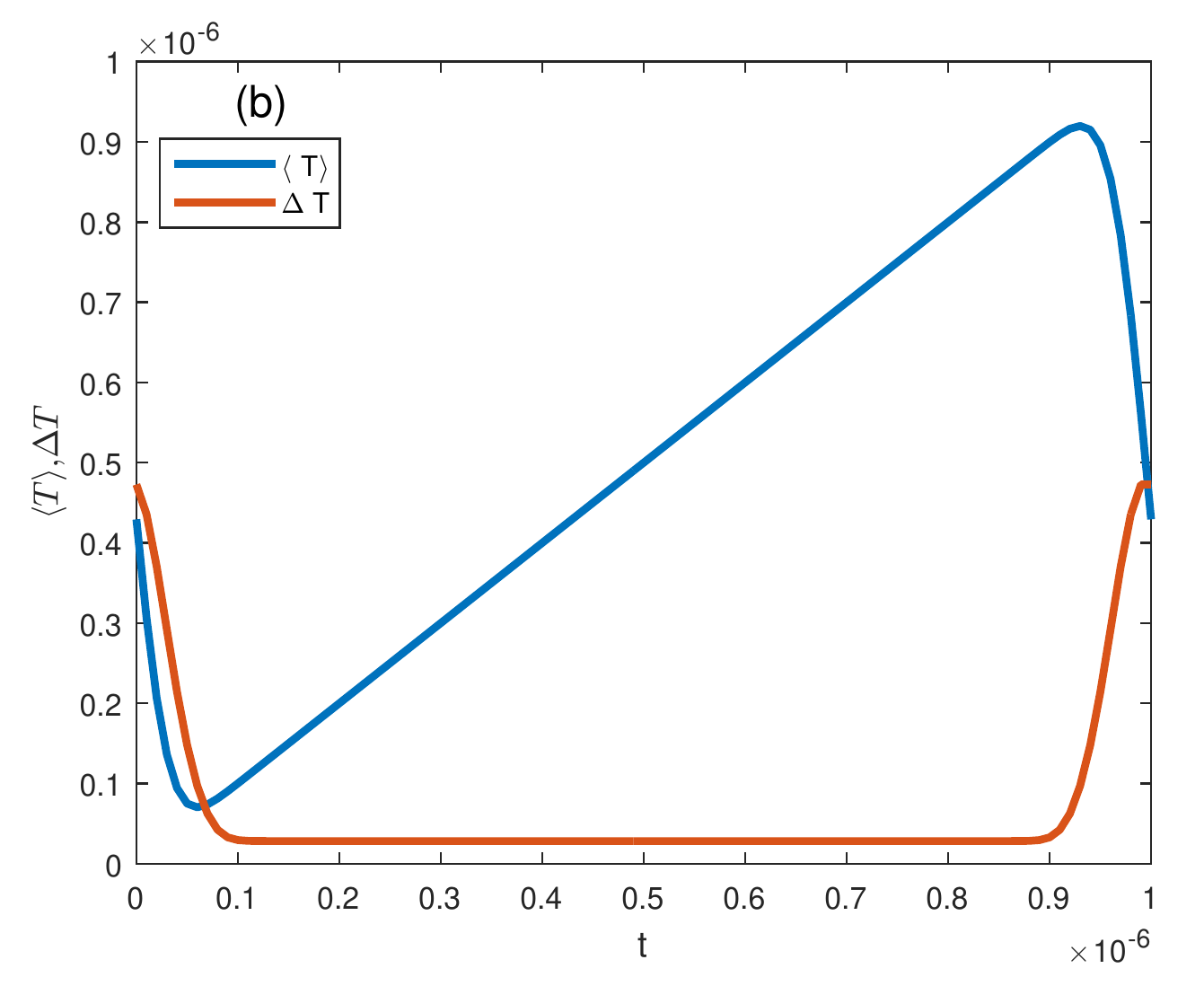}
				\end{tabular}
				\caption{Expectation value $\braket{T_C}$ and deviation $\Delta T_C$ for a SWP time operator over quasi-ideal clock states with $\tau=1\mu s$, $\sigma=\sqrt{d}$, (a) $d=8$ and (b) $d=100$.}
				\label{Fig:3}
			\end{figure}
		\end{center}
	\end{widetext}
	
	
	\section{Wigner's friend scenario with a quasi-ideal clock}
	\label{sec:4}
	
	Now we propose a Wigner's friend scenario. It is constituted of a single lab, inside of which Alice is going to detect the spin of a spin-$\frac{1}{2}$ particle in the $z$ axis. The state is initially prepared in a state
	\begin{equation}
		\ket{+}=\frac{1}{\sqrt 2}(\ket{\uparrow}_S+\ket{\downarrow}_S),
	\end{equation}
	and Alice is going to perform her measurement $\sigma_z^S$ with respect to a classical and ideal clock at time $t_z$. At a further time $t_W$ of her clock, Wigner, who is outside the lab and unaware on how time is passing inside the lab, is going to perform his measurement, a projection over
	\begin{equation}
		\ket{ok}=\cos\left(\frac{\theta}{2}\right)\ket{\uparrow}_S\otimes\ket{\uparrow}_A+e^{i\phi}\sin\left(\frac{\theta}{2}\right)\ket{\downarrow}_S\otimes\ket{\downarrow}_A,
	\end{equation}
	so from Alice's perspective, Wigner's measurement after her measurement will occur with conditional probabilities
	\begin{equation}
		P_A(ok|t_W,\uparrow)=\cos^2\left(\frac{\theta}{2}\right);
	\end{equation}
	\begin{equation}
		P_A(ok|t_W,\downarrow)=\sin^2\left(\frac{\theta}{2}\right).
	\end{equation}
	
	From Wigner's perspective, there is an unitary evolution being carried on inside the lab, and at the time $T_W$ of his SWP clock, he performs his projection. The system is prepared initially in the state
	\begin{equation}
		\rho_{AS}=\ket{\perp}\bra{\perp}_A\otimes\ket{+}\bra{+}_S,
	\end{equation}
	and the clock is prepared in the quasi-ideal clock state
	\begin{equation}
		\rho_C=\ket{\psi(0)}\bra{\psi(0)}.
	\end{equation}
	Since we are equipped with the local asymmetric states only, we must perform a $G$-twirling operation over the product state $\rho=\rho_S\otimes\rho_C$. The global hamiltonian, however, is not complete, for only $H_C$ is known (Eq. (\ref{HC})). We must therefore search for a reasonable hamiltonian capable of describing the evolution
	\begin{equation}\label{vonneumann}
		\ket{+}_S\otimes\ket{\perp}_A\rightarrow\frac{1}{\sqrt 2}(\ket{\uparrow}_S\ket{\uparrow}_A+\ket{\downarrow}_S\ket{\downarrow}_A).
	\end{equation}
	
	For this, we present three possible unitaries. The first one is a instantaneous transition, given by
	\begin{widetext}
		\begin{equation}\label{instant}
			U_t=
			\begin{pmatrix}
				1-\Theta(\Delta t) & 0 & -i\Theta(\Delta t) & 0 & 0 & 0\\
				0 & 1-\Theta(\Delta t) & 0 & 0 & 0 & -i\Theta(\Delta t)\\
				-i\Theta(\Delta t) & 0 & 1-\Theta(\Delta t) & 0 & 0 & 0\\
				0 & 0 & 0 & 1 & 0 & 0\\
				0 & 0 & 0 & 0 & 1 & 0\\
				0 & -i\Theta(\Delta t) & 0 & 0 & 0 & 1-\Theta(\Delta t)
			\end{pmatrix},
		\end{equation}
	\end{widetext}
	where the basis adopted is given by
	\begin{equation}
		\{\ket{\perp}_A\ket{\uparrow(\downarrow)}_S\ket{\uparrow}_A\ket{\uparrow(\downarrow)}_S,\ket{\downarrow}_A\ket{\uparrow(\downarrow)}_S\},
	\end{equation}
	and $\Theta(\Delta t)=\Theta(t-t_z)$ is the Heaviside step function. However, this might be claimed as a too unrealistic description of the evolution inside the lab, and we could also work with the analytical version of the step funcion given by hiperbolic functions
	\begin{widetext}
		\begin{equation}\label{nonperiodic}
			U_t=\frac{1}{\sqrt 2}
			\begin{pmatrix}
				\frac{1-\tanh(\omega_0\Delta t)}{\sqrt{1+\tanh^2(\omega_0\Delta t)}} & 0 & -i\frac{1+\tanh(\omega_0\Delta t)}{\sqrt{1+\tanh^2(\omega_0\Delta t)}} & 0 & 0 & 0\\
				0 & \frac{1-\tanh(\omega_0\Delta t)}{\sqrt{1+\tanh^2(\omega_0\Delta t)}} & 0 & 0 & 0 & -i\frac{1+\tanh(\omega_0\Delta t)}{\sqrt{1+\tanh^2(\omega_0\Delta t)}}\\
				-i\frac{1+\tanh(\omega_0\Delta t)}{\sqrt{1+\tanh^2(\omega_0\Delta t)}} & 0 & \frac{1-\tanh(\omega_0\Delta t)}{\sqrt{1+\tanh^2(\omega_0\Delta t)}} & 0 & 0 & 0\\
				0 & 0 & 0 & \sqrt2 & 0 & 0\\
				0 & 0 & 0 & 0 & \sqrt2 & 0\\
				0 & -i\frac{1+\tanh(\omega_0\Delta t)}{\sqrt{1+\tanh^2(\omega_0\Delta t)}} & 0 & 0 & 0 & \frac{1-\tanh(\omega_0\Delta t)}{\sqrt{1+\tanh^2(\omega_0\Delta t)}}
			\end{pmatrix}.
		\end{equation}
	\end{widetext}
	
	Here, $\omega_0$ is a parameter on how fast the transition happens. Another possible option for the transition is the following, which we are going to call the periodic transition
	
	\begin{widetext}
		\begin{equation}\label{periodic}
			U_t=
			\begin{pmatrix}
				\cos\omega_0t & 0 & -i\sin\omega_0t & 0 & 0 & 0\\
				0 & \cos\omega_0t & 0 & 0 & 0 & -i\sin\omega_0t\\
				-i\sin\omega_0t & 0 & \cos\omega_0t & 0 & 0 & 0\\
				0 & 0 & 0 & 1 & 0 & 0\\
				0 & 0 & 0 & 0 & 1 & 0\\
				0 & -i\sin\omega_0t & 0 & 0 & 0 & \cos\omega_0t
			\end{pmatrix}
		\end{equation}
	\end{widetext}
	where $\omega_0$ again quantifies how fast the transition happens. However, this periodic evolution must be halted at a specific time $T=\left(m+\frac{1}{2}\right)\frac{\pi}{\omega_0}$, with $m\in\mathbb{Z}$, to effectively represent Eq. (\ref{vonneumann}). Otherwise, the measurement is either not completed or is already being undone. This model of periodic transition has already been used in dynamical models for projective measurements \cite{DePasquale2019}, and is going to be adopted as the rulling evolution, since it has a simple generating hamiltonian given by
	\begin{equation}
		H_{AS}=\omega_0(\ket{\perp\uparrow}\bra{\uparrow\uparrow}_{AS}+\ket{\perp\downarrow}\bra{\downarrow\downarrow}_{AS}+h.c.),
	\end{equation}
	
	We are now capable of describing the global hamiltonian, given by
	\begin{equation}
		H=H_{AS}\otimes\mathbb{I}_C+\mathbb{I}_{AS}\otimes H_C,
	\end{equation}
	and thus performing the $G$-twirling over the state
	\begin{equation}
		\rho=\ket{\perp}\bra{\perp}_A\otimes\ket{+}\bra{+}_S\otimes\ket{\psi(0)}\bra{\psi(0)}_C.
	\end{equation}
	
	The relevant quantities in our analysis are going to be given by the differences in probabilities predicted by Alice and Wigner for the outcome $ok$, represented by
	\begin{equation}
		\Delta_0=P_A(ok|t_z,\uparrow)-P_W(ok|T_W),
	\end{equation}
	\begin{equation}
		\Delta_1=P_A(ok|t_z,\downarrow)-P_W(ok|T_W).
	\end{equation}
	The paradox vanishes whenever
	\begin{equation}\label{constr}
		\Delta_0=\Delta_1=0,
	\end{equation}
	a constraint equation we will define as our \emph{consistency condition}. Particularly, if quantum mechanics is a consistent theory, then Eq. (\ref{constr}) should be satisfied for all $\theta\in[0,\pi]$ and $\phi\in[0,2\pi]$ that characterizes Wigner's detection $ok$.
	
	
	\section{Results}\label{sec:5}
	
	\subsection{Satisfying the consistency condition}
	
	It is interesting to see that analytical calculations lead us to two relational density operators for the lab (see Appendix), given approximations ($\sigma\geq \sqrt{d}$ and $d\rightarrow\infty$). The first one is simply a statistical mixture,
	\begin{equation}\label{nonres}
		\rho_{SA}(K)=\frac{1}{2}\ket{\perp}\bra{\perp}_A\otimes\ket{+}\bra{+}_S+\frac{1}{2}\ket{\Phi_+}\bra{\Phi_+},
	\end{equation}
	where $\ket{\Phi_+}=\frac{1}{\sqrt 2}(\ket{\uparrow}_A\ket{\uparrow}_S+\ket{\downarrow}_A\ket{\downarrow}_S)$. However, for specific energies $\omega_0$ of the lab given by 
	\begin{equation}
		\omega_0=\frac{q}{2}\omega, \quad q\in\mathbb{Z},
	\end{equation}
	where $\omega$ is the SWP clock frequency, then the relational state is described as
	\begin{widetext}
		\begin{equation}\label{res}
			\rho_{SA}(K)=\frac{1}{4}
			\begin{pmatrix}
				1+\mathcal{R}(K) & 1+\mathcal{R}(K) & i\mathcal{Q}(K) & 0 & 0 & i\mathcal{Q}(K)\\
				1+\mathcal{R}(K) & 1+\mathcal{R}(K) & i\mathcal{Q}(K) & 0 & 0 & i\mathcal{Q}(K)\\
				-i\mathcal{Q}(K) & -i\mathcal{Q}(K) & 1-\mathcal{R}(K) & 0 & 0 & 1-\mathcal{R}(K)\\
				0 & 0 & 0 & 0 & 0 & 0\\
				0 & 0 & 0 & 0 & 0 & 0\\
				-i\mathcal{Q}(K) & -i\mathcal{Q}(K) & 1-\mathcal{R}(K) & 0 & 0 & 1-\mathcal{R}(K)\\
			\end{pmatrix},
		\end{equation}
	\end{widetext}
	where functions $\mathcal{R}(K)$ and $\mathcal{Q}(K)$ preserve both the periodic behavior of the transition and the proportion $\frac{\omega_0}{\omega}=\frac{q}{2}$ of the resonant behavior. Explicitly, they can be represented as
	\begin{equation}\label{R}
		\mathcal{R}(K)=e^{-\Gamma^2}\frac{\mbox{Re}\left\{\mbox{erf}\left[\frac{\sqrt{2\pi}}{\sigma}\frac{d}{2}+i\Gamma\right]\right\}}{\mbox{erf}\left[\frac{\sqrt{2\pi}}{\sigma}\frac{d}{2}\right]}\cos\left(\frac{2\pi q}{d}K\right),
	\end{equation}
	\begin{equation}\label{Q}
		\mathcal{Q}(K)=e^{-\Gamma^2}\frac{\mbox{Re}\left\{\mbox{erf}\left[\frac{\sqrt{2\pi}}{\sigma}\frac{d}{2}+i\Gamma\right]\right\}}{\mbox{erf}\left[\frac{\sqrt{2\pi}}{\sigma}\frac{d}{2}\right]}\sin\left(\frac{2\pi q}{d}K\right),
	\end{equation}
	with
	\begin{equation}
		\Gamma=\sqrt{2\pi}\frac{\sigma}{d}\frac{\omega_0}{\omega}=\sqrt{\frac{\pi}{2}}\frac{\sigma}{d}q,
	\end{equation}
	$\mbox{erf}[x]$ being the error function and $K$ being related to the time detected by Wigner in his clock, $K=\frac{T d}{\tau}$.
	
	According to what was discussed in Sec. \ref{sec:4}, Wigner must perform his measurement at a time $T_W=\left(m+\frac{1}{2}\right)\frac{\pi}{\omega_0}$ in order to detect the von Neumann measurement inside the lab. This time is related to a pointer state $K_W=\frac{T_W\,d}{\tau}=\left(m+\frac{1}{2}\right)\frac{d\,\pi}{\omega_0\,\tau}$, and for resonant evolution, 
	\begin{equation}
		K_W=\left(m+\frac{1}{2}\right)\frac{d}{q}, \quad m,q\in\mathbb{Z},
	\end{equation}
	which can explain the source of the resonant behavior. If Wigner is suposed to halt the evolution at a time $T_W$, this time must be an eigenvalue of his clock, i.e., it must be related to a pointer $\ket{\theta_K}$ of his clock. But since these pointers can be associated only with integer (for even $d$) or half-integer (for odd $d$) numbers, by definition, then the transition must be characterized by a resonant frequency. If not, then Wigner is not allowed to perform its measurement at the proper instant, and will always do it over a state in which the measurement has not yet been completed or is already being undone, resulting in the mixed relative state given by Eq. (\ref{nonres}).
	
	At the specific pointer $K_W$, the resonant relative state is given by
	\begin{widetext}
		\begin{equation}
			\rho_{SA}(K_W)=\frac14
			\begin{pmatrix}
				1-\mathcal{R}(0) & 1-\mathcal{R}(0) & 0 & 0 & 0 & 0\\
				1-\mathcal{R}(0) & 1-\mathcal{R}(0) & 0 & 0 & 0 & 0\\
				0 & 0 & 1+\mathcal{R}(0) & 0 & 0 & 1+\mathcal{R}(0)\\
				0 & 0 & 0 & 0 & 0 & 0\\
				0 & 0 & 0 & 0 & 0 & 0\\
				0 & 0 & 1+\mathcal{R}(0) & 0 & 0 & 1+\mathcal{R}(0)
			\end{pmatrix},
		\end{equation}
	\end{widetext}
	since $\mathcal{Q}(K_W)=0$ and $\mathcal{R}(K_W)=-\mathcal{R}(0)$. From this, it is possible to obtain Wigner's conditional probability $P_W(ok|T_W)=\mbox{Tr}\{\ket{ok}\bra{ok}\rho_{SA}(K_W)\}$, leading to differences in the prediction given by
	\begin{widetext}
		\begin{equation}
			\Delta_{0(1)}=\frac{1\pm\cos\theta}{2}-\frac14\left(1+e^{-\left(q\sqrt{\frac{\pi}{2}}\frac{\sigma}{d}\right)^2}\frac{\mbox{Re}\left\{\mbox{erf}\left[\sqrt{\frac{\pi}{2}}\frac{d}{\sigma}+iq\sqrt{\frac{\pi}{2}}\frac{\sigma}{d}\right]\right\}}{\mbox{erf}\left[\sqrt{\frac{\pi}{2}}\frac{d}{\sigma}\right]}\right)(1+\sin\theta\cos\phi).
		\end{equation}
	\end{widetext}
	
	In Fig.\ref{Fig:4} it is possible to see the values of $\theta$ and $\phi$ for which both $\Delta_0$ and $\Delta_1$ are null. The consistency condition (Eq. (\ref{constr})) is satisfied whenever a red and a black line cross (blue dots). The paradox evidently vanishes for very specific measurements $\ket{ok}$ performed by Wigner. Notice that, for $\frac{\sigma}{d}\rightarrow0$ (within the restrictions needed for our calculations to be valid, i.e., $\sigma\geq \sqrt{d}$), the only values of $\theta$ and $\phi$ for which the consistency condition is satisfied are related to the observables
	\begin{equation}
		\ket{ok}=\frac{1}{\sqrt2}(\ket{\uparrow}_A\ket{\uparrow}_S\pm i\ket{\downarrow}_A\ket{\downarrow}_S),
	\end{equation}
	precisely the same observables that rule out the paradox for the original Wigner's friend scenario introduced in Sec. \ref{sec:2}. It is interesting to see that for a clock state with gaussian spread $\sigma=\sqrt{d}$ (called a \emph{symmetric state}), if $d$ is large enough, it is possible to recover the same scenario that the one associated to a shared classical clock.
	
	For a ratio $\frac{\sigma}{d}\rightarrow1$, otherwise, the relative state becomes closer to the mixed relative state given by equation Eq. (\ref{nonres}), and the only observable that Wigner can measure without raising a paradox approaches from
	\begin{equation}
		\ket{ok}=\frac{1}{\sqrt2}(\ket{\uparrow}_A\ket{\uparrow}_S+\ket{\downarrow}_A\ket{\downarrow}_S)=\ket{\Phi_+}.
	\end{equation}
	It is convenient to know that the uncertainty on the clock state preparation can control which observable is allowed to be done, and even if this model is not capable of ruling out the paradox for any observable $ok$, this might indicate that the right choice of evolution inside the lab associated to the quasi-ideal clock states could eventually catalyze the reduced state observed by Alice.
	\begin{widetext}
		\begin{center}
			\begin{figure}[hbt!]
				\centering
				\begin{tabular}{|c|c|}\hline
					\includegraphics[scale=0.5]{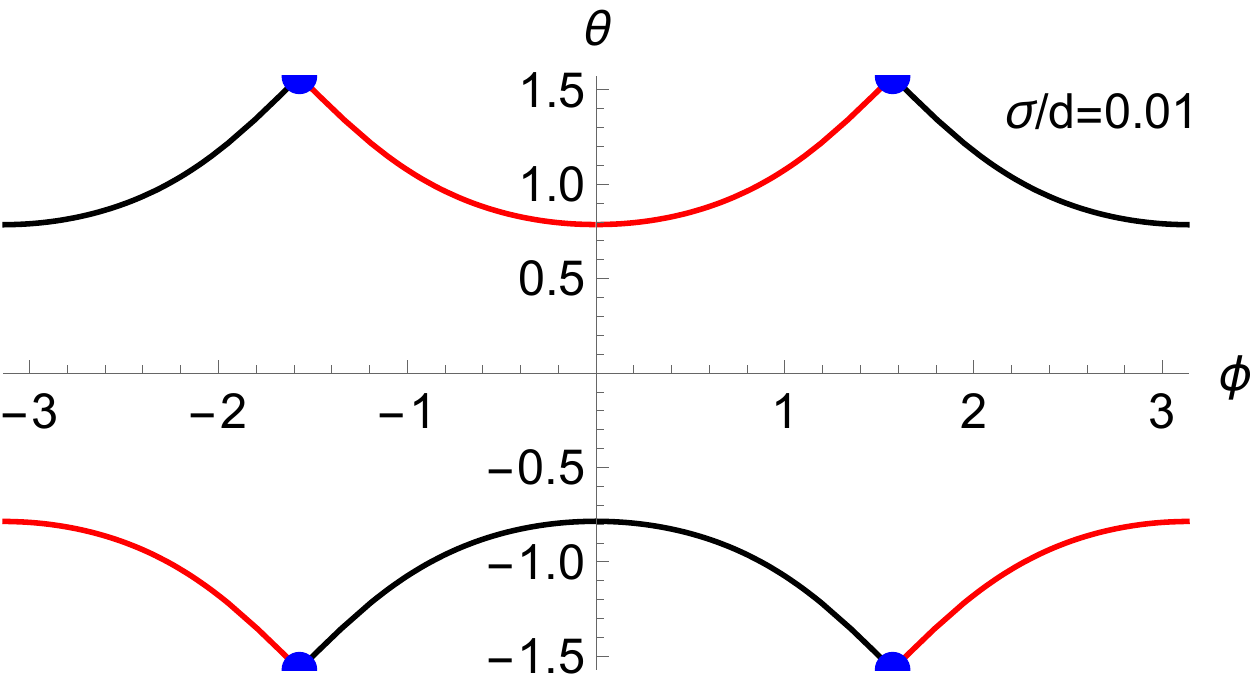} & \includegraphics[scale=0.5]{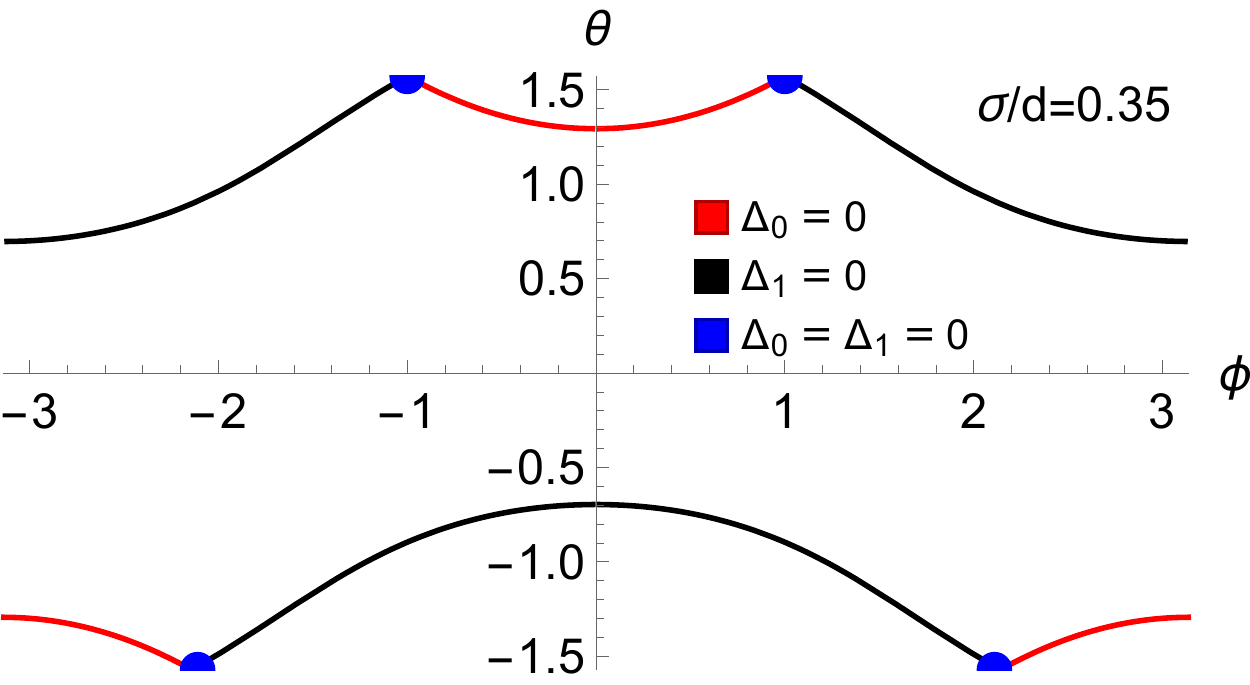}\\\hline
					\includegraphics[scale=0.5]{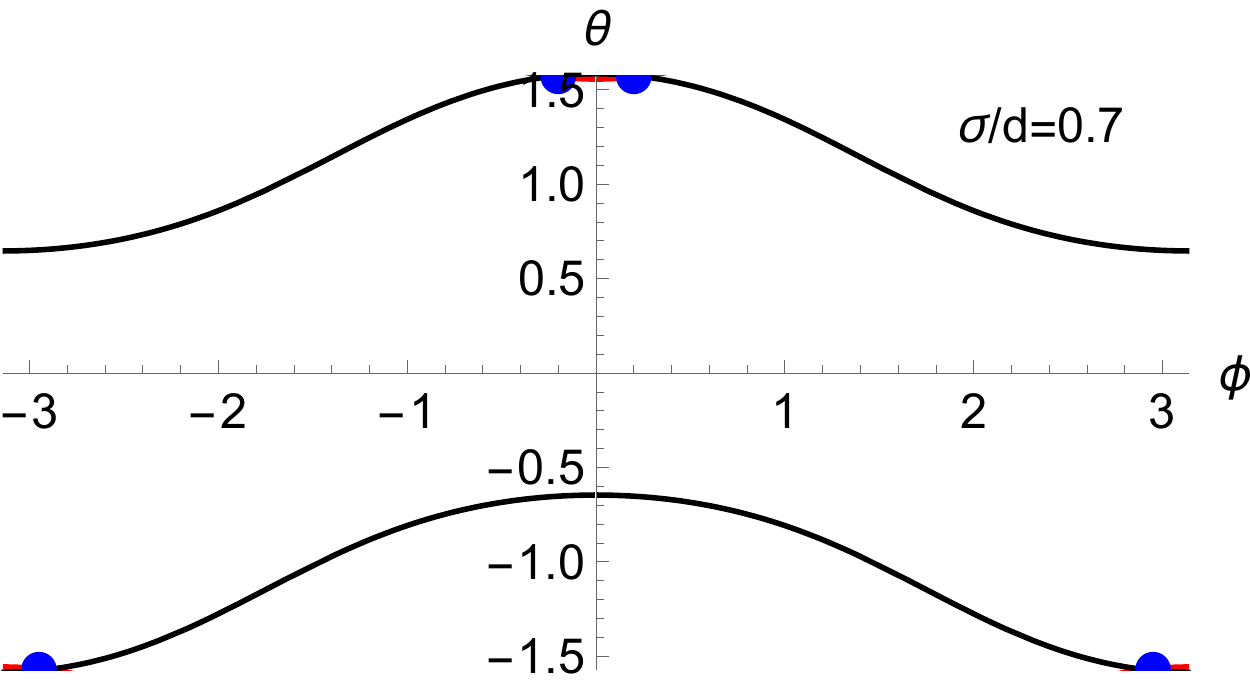} & \includegraphics[scale=0.5]{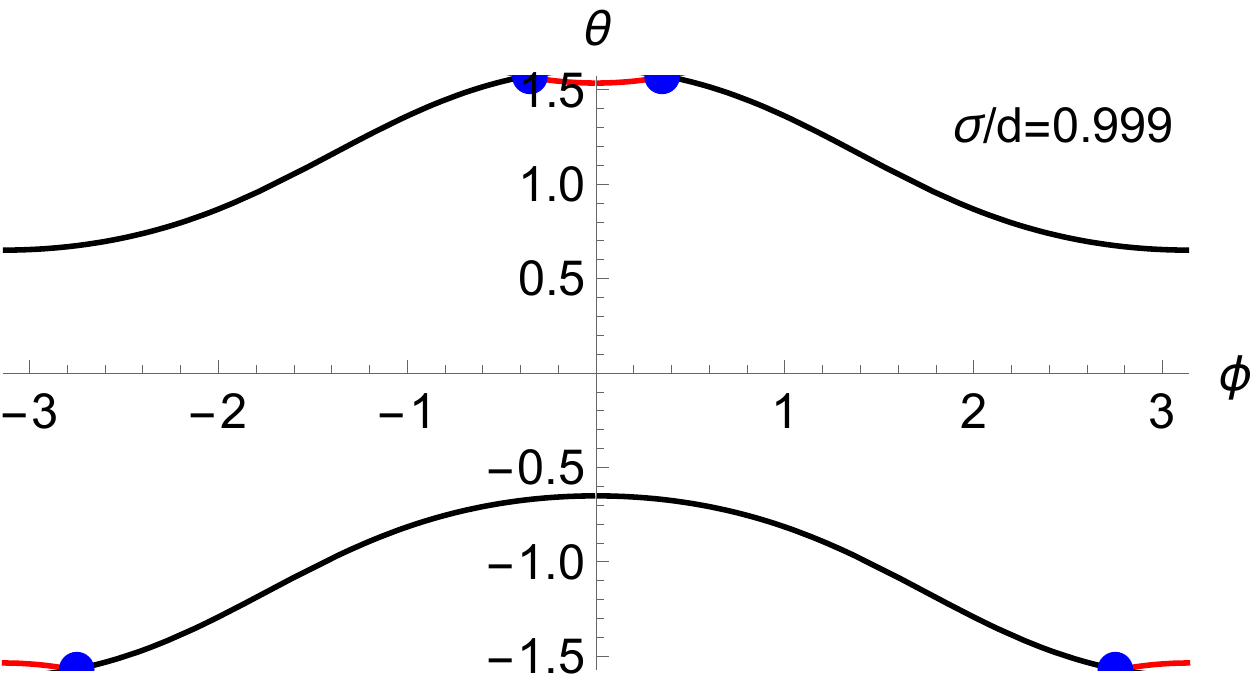}\\\hline
				\end{tabular}
				\caption{Values of $\theta$ and $\phi$ of Wigner's projection $\ket{ok}$, for which $\Delta_0$ (red lines) and $\Delta_1$ (black lines) are null, for different ratios $\frac{\sigma}{d}$, $d\gg1$, $\omega_0=\frac12\omega$. Blue dots refer to points in which the consistency condition is satisfied.}
				\label{Fig:4}
			\end{figure}
		\end{center}
	\end{widetext}
	
	
	\subsection{Analysis of Wigner's measurements}
	
	From a resource-theoretic point of view, it is important to analyse which measurements are allowed for Wigner. If time is being internalized in the sense that Wigner should no longer be aware of how time is passing inside the lab, his measurements might be such that no resource is generated (in this case, asymmetry with respect to time evolution). For that purpose, we ought to look at the invariant subspaces of $\mathcal{H}=\mathcal{H}_{SA}\otimes\mathcal{H}_C$ with respect to the global time evolution, the so-called \emph{charge sectors} \cite{Bartlett2007, Carmo2020}.
	
	The clock hamiltonian is already diagonalized, and by diagonalizing the lab hamiltonian, we can work in the basis $\{\ket{-\omega_{0{\uparrow(\downarrow)}},n},\ket{0_{{\uparrow(\downarrow)}},n},\ket{+\omega_0{{\uparrow(\downarrow)}},n}\}$, where 
	
	\begin{equation}
		\ket{\pm\omega_{0\uparrow}}_{SA}=\frac{1}{\sqrt2}(\ket{\perp\uparrow}_{SA}\pm\ket{\uparrow\uparrow}_{SA}),
	\end{equation}
	\begin{equation}
		\ket{\pm\omega_{0\downarrow}}_{SA}=\frac{1}{\sqrt2}(\ket{\perp\downarrow}_{SA}\pm\ket{\downarrow\downarrow}_{SA}),
	\end{equation}
	\begin{equation}
		\ket{0_\uparrow}_{SA}=\ket{\uparrow\downarrow}_{SA}, \quad \ket{0_\downarrow}_{SA}=\ket{\downarrow\uparrow}_{SA},
	\end{equation}
	are eigenstates associated to the eigenvalues $\{\pm\omega_0,0\}$ of the lab hamiltonian.
	
	The charge sectors are constituted of subspaces of $\mathcal{H}$ associated to the same eigenvalues of the global hamiltonian $H$. In a first moment, each charge sector will have dimension 2, associated to the numbers $n\omega-\omega_0$, $n\omega$ or $n\omega+\omega_0$, for $n\in[0,d-1]$. However, when $\omega_0=\frac{q}{2}\omega$, $q\in\mathbb{Z}$, some of these eigenvalues coincide with each other, ensuring nontrivial charge sectors of higher dimension.
	
	For odd $q$, such as the one adopted in Fig. \ref{Fig:4}, the charge sectors associated to $n\omega$ remain 2-dimensional, while the one associated to $n\omega+\omega_0$ is 4-dimensional and generated by the basis
	\begin{equation}
		\{\ket{+\omega_{0{\uparrow(\downarrow)}},n},\ket{-\omega_{0{\uparrow(\downarrow)}},n+q}\},
	\end{equation}
	for $0\leq|n+q|\leq d-1$. For even $q$, the nontrivial charge sector associated to this number is 6-dimensional and generated by the basis
	\begin{equation}
		\{\ket{+\omega_{0{\uparrow(\downarrow)}},n},\ket{0_{\uparrow(\downarrow)},n+q},\ket{-\omega_{0{\uparrow(\downarrow)}},n+2q}\},
	\end{equation}
	for $0\leq|n+2q|\leq d-1$. We can see then why the resonance emerges, since it is the only regime in which nontrivial charge sectors occur. Larger charge sectors can protect more information against the dephasing provoked by the $G$-twirling operation \cite{Carmo2020}.
	
	If Wigner is allowed to perform only symmetric operations, i.e., operations that preserve the asymmetry of the global state, then he should only perform measurements that cannot transfer information from one charge sector to another \cite{Gour2008, Chitambar2019}. For our definition of $\ket{ok}$, the projector $\Pi_{ok}$ is written in the diagonal lab basis as
	\begin{equation}
		\Pi_{ok}=\frac{1}{2}
		\begin{pmatrix}
			A & O & -A\\
			O & O & O\\
			-A & O & A
		\end{pmatrix},
	\end{equation}
	where 
	\begin{equation}
		A=
		\begin{pmatrix}
			\cos^2\frac{\theta}{2} & \frac{1}{2} e^{-i\phi}\sin\theta\\
			\frac{1}{2} e^{i\phi}\sin\theta & \sin^2\frac{\theta}{2}
		\end{pmatrix},
	\end{equation}
	and $O$ is a $2\times2$ null matrix. Writing down then the operator $\Pi_{ok}\otimes\mathbb{I}_C$ in the diagonal global basis, e.g. for $d=3$ and $q=1$, we are led to
	\begin{widetext}
		\begin{equation}
			\Pi_{ok}\otimes\mathbb{I}_C=\frac{1}{2}
			\begin{tikzpicture}[baseline]
				\matrix (m)[
				matrix of math nodes,
				left delimiter=(, right delimiter=),
				nodes in empty cells,
				minimum width=width("100"),
				] {
					A & O & -A &    &   &    &    &   &    \\
					O & O &  O &    &   &    &    &   &    \\
					-A & O &  A &    &   &    &    &   &    \\
					&   &    &  A & O & -A &    &   &    \\
					&   &    &  O & O &  O &    &   &    \\
					&   &    & -A & O &  A &    &   &    \\
					&   &    &    &   &    &  A & O & -A \\
					&   &    &    &   &    &  O & O &  O \\
					&   &    &    &   &    & -A & O &  A \\
				} ;
				\draw(m-1-1.south west) rectangle (m-1-1.north east);
				\draw(m-2-2.south west) rectangle (m-2-2.north east);
				\draw (m-3-3.north west) rectangle (m-4-4.south east);
				\draw (m-5-5.north west) rectangle (m-5-5.south east);
				\draw (m-6-6.north west) rectangle (m-7-7.south east);
				\draw (m-8-8.north west) rectangle (m-8-8.south east);
				\draw (m-9-9.north west) rectangle (m-9-9.south east);
			\end{tikzpicture},
		\end{equation}
	\end{widetext}
	which clearly allows for information flow between charge sectors (black rectangles in the matrix). However, if we impose this restriction over Wigner measurements, then he would be also forbidden of measuring time, since the projector over a pointer state, given in this basis as
	\begin{equation}
		\mathbb{I}_{SA}\otimes\Pi_K^C=\mathbb{I}_{SA}\otimes\frac{1}{d}\sum_{n,n'=0}^de^{-i2\pi(n-n')k/d}\ket{n}\bra{n'},
	\end{equation}
	also allows for information flow between charge sectors. There is thus no point on restricting Wigner's measurement to free operations only, since he would then be forbidden of consulting his clock. It is crucial to consider, however, that no asymmetry monotone can work in such a scenario, given that Wigner is free to generate resource at his will.
	
	One might further ask if the consistency condition would be satisfied for time-symmetric local operations on the lab, even with asymmetric operations performed locally on the clock. Such operations are convex combinations of the projectors
	\begin{equation}
		\ket{ok_{sym}^\pm}=\frac{1}{\sqrt2}(\ket{\xi}_S\otimes\ket{\perp}_A\pm\ket{ok}_{SA}), 
	\end{equation}
	where
	\begin{equation}
		\ket{\xi}_S=\cos\frac{\theta}{2}\ket{\uparrow}_S+e^{i\phi}\sin\frac{\theta}{2}\ket{\downarrow}_S.
	\end{equation}
	Even for such measurements, it is easy to see that the consistency condition is not satisfied for all $(\theta,\phi)$. If Wigner is performing a projection over one of the two $ok_{sym}$, then 
	\begin{equation}
		\Delta_{0(1)}=\frac14\left[\frac12\pm\cos\theta-\cos\phi\sin\theta\right]
	\end{equation}
	and the uncertainty $\sigma$ of the quasi-ideal clock state plays no role on the statistics of these symmetric measurements. Wigner agrees with Alice's predictions only when $(\theta,\phi)=(\pm\frac{\pi}{2},\pm\frac{\pi}{3})$, and the paradox thus persists.
	
	
	\section{Conclusions} \label{sec:6}
	
	This work had the aim to study the consequences of the insertion of a feasible quantum clock in a Wigner's friend scenario. By working with a Salecker-Wigner-Peres clock \cite{Salecker1958, Peres1980} and Woods, Silva and Oppenheim's quasi-ideal clock states \cite{Woods2019} it was possible to internalize time in a WFS in a Page-Wootters formalism. The choice of a periodic model of transition to govern the lab dynamics \cite{DePasquale2019} led to interesting results, indicating that the paradox still does not vanish for any measurement made by Wigner, bur rather for very specific observables controled by the ratios $\frac{\omega_0}{\omega}$ between the lab dynamics and the SWP clock frequency, and $\frac{\sigma}{d}$ of the clock states's gaussian spread. An analysis of how the process of time internalization affects Wigner's possible measurements indicates that restricting them to symmetric operations implies in forbidding Wigner to consult his clock. Furthermore, the paradox persists even for time symmetric operations over the lab.
	
	This result might imply that the quasi-ideal clock state, with its intrinsic uncertainty $\sigma$, is not enough to trigger the desired decoherent behavior that Alice observes when performing her measurement. Indeed, decoherence can be considered as part of the definition of a measurement \cite{Zukowski2020}, and the insertion of uncontrolled degrees of freedom might be unavoidable.
	
	Other possible claim is that the consistency condition imposed by the constraint in equation Eq. (\ref{constr}) is unnecessary. Since Alice and Wigner have access to different parts of the global state, it is not reasonable to demand that they should predict the same probability distributions. The subjectivity of objective measurements due to different reference frames has recently been shown \cite{Le2020}.
	
	Monogamy between Alice and the spin-$\frac{1}{2}$ particle might be preventing the clock of fully accessing the lab dynamics, and thus stealing any coherence \cite{Costa2014,Leggio2015}. There is also a theorem ensuring the possibility of a catalytic conversion from the entangled state to the reduced state through the insertion of a clock \cite{Marvian2016}.
	
	Finally, there is the need of testing other models for the lab dynamics. Particularly, investigating the entanglement in the so-called \emph{internal states} \cite{Carmo2020}, and how quantifiers as shared asymmetry \cite{Carmo2020, Martinelli2019, Mendes2019} would act in this sort of scenario could cast a light on this fundamental problem.\\
	
	\section*{Acknowledgements}
	
	The authors would like to thank Tiago Martinelli, Rafael S.
	Carmo, Eduardo Duzzioni, and Rafael Rabelo for the helpful discussions and valuable comments. The project was
	funded by Brazilian funding agencies CNPq (Grant No.
	307028/2019-4), and FAPESP (Grant No. 2017/03727-0).
	The authors also acknowledge financial support in part by
	the Coordenação de Aperfeiçoamento de Pessoal de Nível
	Superior - Brasil (CAPES) (Finance Code 001) and by the
	Brazilian National Institute for Science and Technology of
	Quantum Information [CNPq INCT-IQ (465469/2014-0)].
	
	
	
	
	\appendix
	
	\section{$G$-twirling over the global state}
	
	The non-interacting hamiltonian generates a global evolution given by
	\begin{equation}
		U_t=e^{-iH_{AS}t}\otimes e^{-iH_Ct}.
	\end{equation}
	The $G$-twriling operation will therefore be an integration whose integrand is the state
	\begin{equation}
		U_t\rho U_t^\dagger=\rho_{SA}(t)\otimes\rho_C(t).
	\end{equation}
	The time-evolved lab state is explicitly given by
	\begin{widetext}
		\begin{equation}
			\rho_{AS}(t)=\frac12
			\begin{pmatrix}
				\cos^2\omega_0t & \cos^2\omega_0t & \frac{i}{2}\sin2\omega_0t & 0 & 0 & \frac{i}{2}\sin2\omega_0t\\
				\cos^2\omega_0t & \cos^2\omega_0t & \frac{i}{2}\sin2\omega_0t & 0 & 0 & \frac{i}{2}\sin2\omega_0t\\
				-\frac{i}{2}\sin2\omega_0t & -\frac{i}{2}\sin2\omega_0t & \sin^2\omega_0t & 0 & 0 & \sin^2\omega_0t\\
				0 & 0 & 0 & 0 & 0 & 0\\
				0 & 0 & 0 & 0 & 0 & 0\\
				\frac{i}{2}\sin2\omega_0t & -\frac{i}{2}\sin2\omega_0t & \sin^2\omega_0t & 0 & 0 & \sin^2\omega_0t
			\end{pmatrix},
		\end{equation}
	\end{widetext}
	while for $d\rightarrow\infty$, the time-evolved clock state will be given by
	\begin{equation}\label{clock1}
		\rho_C(t)\approx\ket{\psi(td/\tau)}\bra{\psi(td/\tau)},
	\end{equation}
	up to an exponentially vanishing error. We can therefore work with integrations of the terms of $U_t\rho U_t^\dagger$, that will be products of one of three basis functions $1$, $e^{\pm i2\omega_0t}$ and the time-dependent term of Eq. \ref{clock1}. Explicitly, it can be written as
	\begin{widetext}
		\begin{equation}
			\rho_C(t)\approx\sum_{k,k'\in S_d(td/\tau)}|A|^2e^{-\frac{\pi}{\sigma^2}(k-td/\tau)^2}e^{-\frac{\pi}{\sigma^2}(k'-td/\tau)^2}e^{i2\pi n_0(k-k')/d}\ket{\theta_k}\bra{\theta_{k'}}. 
		\end{equation}
	\end{widetext}
	
	A result by Woods, Silva and Oppenheim \cite{Woods2019} ensures that, if $\sigma\geq\sqrt{d}$, then $|A|$ is nearly constant in time, and within this range of $\sigma$ we are allowed to take the normalizing factor out of the integral. The summation limits, however, still depend on time, and in the first moment there is no way the integral and the summation commute.
	
	However, an analysis of the behavior of $S_d(td/\tau)$ with respect to $t$ leads to the result
	\begin{equation}
		S_d(td/\tau)=S_d(n), \quad t\in\left(\frac{\tau}{d}(n-1),\frac{\tau}{d}n\right], n\in\mathbb{Z},
	\end{equation}
	which allows us to write the $G$-twirling operation as
	\begin{widetext}
		\begin{equation}
			\begin{split}
				\mathcal{G}
				\begin{bmatrix}
					1\\
					e^{\pm i2\omega_0t}
				\end{bmatrix}
				=\lim_{N\rightarrow\infty}\frac{|A|^2\sigma d}{2\sqrt2\tau N}\sum_{n=-N}^N\sum_{k,k'\in S_d(n)}e^{-\frac{\pi}{2\sigma^2}(k-k')^2}e^{i2\pi n_0(k-k')/d}\ket{\theta_k}\bra{\theta_{k'}}\\\times\int_{\frac{\tau}{d}(n-1)}^{\frac{\tau}{d}n}e^{-\frac{\pi}{\sigma^2}\left(\frac{\sqrt2d}{\tau}t-\frac{k+k'}{\sqrt2}\right)^2}
				\begin{pmatrix}
					1\\
					e^{\pm i2\omega_0t}
				\end{pmatrix}dt.
			\end{split}
		\end{equation}
	\end{widetext}
	
	For the basic function $1$, this integration leads to
	
	\begin{widetext}
		\begin{equation}
			\begin{split}
				\mathcal{G}[1]=\lim_{N\rightarrow\infty}\frac{|A|^2\sigma^2}{8N}\sum_{n=-N}^N\sum_{k,k'\in S_d(n)}\ket{\theta_k}\bra{\theta_{k'}}e^{-\frac{\pi}{2\sigma^2}(k-k')^2}e^{i2\pi n_0(k-k')/d}\\\times\left\{\mbox{erf}\left[\frac{\sqrt{2\pi}}{\sigma}\left(n-\frac{k+k'}{2}\right)\right]-\mbox{erf}\left[\frac{\sqrt{2\pi}}{\sigma}\left(n-1-\frac{k+k'}{2}\right)\right]\right\};
			\end{split}
		\end{equation}
	\end{widetext}
	while for $e^{\pm i2\omega_0t}$, it leads to
	\begin{widetext}
		\begin{equation}
			\begin{split}
				\mathcal{G}[e^{\pm i2\omega_0t}]=\lim_{N\rightarrow\infty}\frac{|A|^2\sigma e^{-\Gamma^2}}{4\sqrt2d}\sum_{n=-N}^N\sum_{k,k'\in S_d(n)}\ket{\theta_k}\bra{\theta_{k'}}e^{-\frac{\pi}{2\sigma^2}(k-k')^2}e^{i2\pi n_0(k-k')/d}e^{\pm i\frac{\sqrt{2\pi}}{\sigma}\Gamma(k+k')}\\\times\left\{\mbox{erf}\left[\frac{\sqrt{2\pi}}{\sigma}\left(n-\frac{k+k'}{2}\right)\pm i\Gamma\right]-\mbox{erf}\left[\frac{\sqrt{2\pi}}{\sigma}\left(n-1-\frac{k+k'}{2}\right)\pm i\Gamma\right]\right\}.
			\end{split}
		\end{equation}
	\end{widetext}
	
	Since we are not interest directly on the global symmetric state $\mathcal{G}[\rho]$, but rather in the relational state given by Eq. (\ref{relative}), we can start to project these results over a specific pointer state $\ket{\theta_K}$. One must remember, however, to take in account not only the term associated to $\ket{\theta_K}$, but every other element corresponding to $\ket{\theta_{K+md}}$, $m\in\mathbb{Z}$, since the pointer basis is infinitely degenerated. For $\mathcal{G}[1]$ and $\mathcal{G}[e^{\pm i2\omega_0t}$, this analyses lead to
	\begin{widetext}
		\begin{equation}
			\Pi_K^C\mathcal{G}[1]\Pi_K^C=\lim_{N\rightarrow\infty}\frac{|A|^2\sigma^2}{8N}\frac{2N}{d}2\mbox{erf}\left[\frac{\sqrt{2\pi}}{\sigma}\frac{d}{2}\right]=\frac{|A|^2\sigma^2}{2d}\mbox{erf}\left[\frac{\sqrt{2\pi}}{\sigma}\frac{d}{2}\right];
		\end{equation}
		
		\begin{equation}\label{exp1}
			\begin{split}
				\Pi_K^C\mathcal{G}[e^{\pm i2\omega_0t}]\Pi_K^C=\lim_{N\rightarrow\infty}\frac{|A|^2\sigma^2 e^{-\Gamma^2}}{4N}e^{\pm i2\frac{\sqrt{2\pi}}{\sigma}\Gamma K}\mbox{Re}\left\{\mbox{erf}\left[\frac{\sqrt{2\pi}}{\sigma}\frac{d}{2}\pm i\Gamma\right]\right\}\sum_{m=-N/d}^{N/d}e^{\pm i2\frac{\sqrt{2\pi}}{\sigma}\Gamma md}.
			\end{split}
		\end{equation}
	\end{widetext}
	
	Notice that the limit in Eq. (\ref{exp1}) goes to $0$ unless each term in the summation is equal to $1$. This will happen whenever
	\begin{equation}
		\frac{\sqrt{2\pi}}{\sigma}\Gamma d=q\pi\iff\omega_0\tau=q\pi, \quad q\in\mathbb{Z},
	\end{equation}
	which means that 
	\begin{equation}
		\frac{\omega_0}{\omega}=\frac{q}{2}, \quad q\in\mathbb{Z}.
	\end{equation}
	In this case, we can rewrite Eq. (\ref{exp1}) as
	\begin{widetext}
		\begin{equation}
			\Pi_K^C\mathcal{G}[e^{\pm i2\omega_0t}]\Pi_K^C=
			\begin{cases}
				\frac{|A|^2\sigma^2 e^{-\Gamma^2}}{2d}e^{\pm i2\frac{\sqrt{2\pi}}{\sigma}\Gamma K}\mbox{Re}\left\{\mbox{erf}\left[\frac{\sqrt{2\pi}}{\sigma}\frac{d}{2}\pm i\Gamma\right]\right\}, \quad\mbox{ if }\omega_0=\frac{q}{2}\omega;\\
				0,\mbox{ otherwise}
			\end{cases}.
		\end{equation}
	\end{widetext}
	
	Therefore, we can finally describe every entry of the relative lab state as a linear combination of the symmetric functions $\mathcal{G}[1]$ and $\mathcal{G}[e^{\pm i2\omega_0t}]$, such that
	\begin{widetext}
		\begin{equation}
			\Pi_K^C\mathcal{G}[\cos^2\omega_0t]\Pi_K^C=\frac12\Pi_K^C\mathcal{G}[1]\Pi_K^C+\frac14(\Pi_K^C\mathcal{G}[e^{i2\omega_0t}]\Pi_K^C+\Pi_K^C\mathcal{G}[e^{-i2\omega_0t}]\Pi_K^C);
		\end{equation}
		\begin{equation}
			\Pi_K^C\mathcal{G}[\sin^2\omega_0t]\Pi_K^C=\frac12\Pi_K^C\mathcal{G}[1]\Pi_K^C-\frac14(\Pi_K^C\mathcal{G}[e^{i2\omega_0t}]\Pi_K^C+\Pi_K^C\mathcal{G}[e^{-i2\omega_0t}]\Pi_K^C);
		\end{equation}
		\begin{equation}
			\Pi_K^C\mathcal{G}[\sin2\omega_0t]\Pi_K^C=\frac{1}{2i}(\Pi_K^C\mathcal{G}[e^{i2\omega_0t}]\Pi_K^C-\Pi_K^C\mathcal{G}[e^{-i2\omega_0t}]\Pi_K^C),
		\end{equation}
	\end{widetext}
	finally obtaining the relative state
	\begin{widetext}
		\begin{equation}\label{mixed}
			\rho_{FS}^W(K)=\mbox{Tr}_C\left\{\frac{(\mathbb{I}_{FS}\otimes\Pi_K^C)\mathcal{G}[\rho](\mathbb{I}_{FS}\otimes\Pi_K^C)}{\mbox{Tr}\{(\mathbb{I}_{FS}\otimes\Pi_K^C)\mathcal{G}[\rho]\}}\right\}=\frac14
			\begin{pmatrix}
				1 & 1 & 0 & 0 & 0 & 0\\
				1 & 1 & 0 & 0 & 0 & 0\\
				0 & 0 & 1 & 0 & 0 & 1\\
				0 & 0 & 0 & 0 & 0 & 0\\
				0 & 0 & 0 & 0 & 0 & 0\\
				0 & 0 & 1 & 0 & 0 & 1
			\end{pmatrix},
		\end{equation}
	\end{widetext}
	if $\frac{\omega_0}{\omega}\neq\frac{q}{2}$, or
	\begin{widetext}
		\begin{equation}
			\rho(K)=\frac{1}{4}
			\begin{pmatrix}
				1+\mathcal{R}(K) & 1+\mathcal{R}(K) & i\mathcal{Q}(K) & 0 & 0 & i\mathcal{Q}(K)\\
				1+\mathcal{R}(K) & 1+\mathcal{R}(K) & i\mathcal{Q}(K) & 0 & 0 & i\mathcal{Q}(K)\\
				-i\mathcal{Q}(K) & -i\mathcal{Q}(K) & 1-\mathcal{R}(K) & 0 & 0 & 1-\mathcal{R}(K)\\
				0 & 0 & 0 & 0 & 0 & 0\\
				0 & 0 & 0 & 0 & 0 & 0\\
				-i\mathcal{Q}(K) & -i\mathcal{Q}(K) & 1-\mathcal{R}(K) & 0 & 0 & 1-\mathcal{R}(K)\\
			\end{pmatrix},
		\end{equation}
	\end{widetext}
	if $\frac{\omega_0}{\omega}=\frac{q}{2}$, with $\mathcal{R}(K)$ and $\mathcal{Q}(K)$ already defined by Eqs. (\ref{R}) and (\ref{Q}).
	
	
	
\end{document}